# MALICIOUS GENAI CHROME EXTENSIONS: UNPACKING DATA EXFILTRATION AND MALICIOUS BEHAVIOURS

Shresta B.Seetharam, Mohamed Nabeel & William Melicher

*Palo Alto Networks, USA*

sseetharam@paloaltonetworks.com
mmohamednabe@paloaltonetworks.com
bmelicher@paloaltonetworks.com

## ABSTRACT

The rapid proliferation of AI and GenAI tools has extended to the *Chrome Web Store*. Cybercriminals are exploiting this trend, deploying malicious *Chrome* extensions posing as AI tools or impersonating popular GenAI models to target users. These extensions often appear legitimate while secretly exfiltrating sensitive data or redirecting users' web traffic to attacker-controlled domains.

To examine the impact of this trend on the browser extension ecosystem, we curated a dataset of 5,551 AI-themed extensions released over a nine-month period to the *Chrome Web Store*. Using a multi-signal detection methodology that combines manifest analysis, domain reputation, and runtime network behaviour, supplemented with human review, we identified 154 previously undetected malicious *Chrome* extensions. Together with extensions known from public threat research disclosures, this resulted in a final set of 341 malicious extensions for analysis. Of these, 29 were GenAI-related, forming the focus of our in-depth analysis and disclosure.

We deconstruct representative GenAI cases, including Supersonic AI, DeepSeek AI | Free AI Assistant, and Perplexity Search, to illustrate attacker techniques such as adversary-in-the-browser, impersonation, bait-and-switch updates, query hijacking, and redirection. Our findings show that threat actors are leveraging GenAI trends and exploiting browser extension APIs and settings for malicious purposes. This demonstrates that the browser extension threat landscape is directly evolving alongside the rapid adoption of GenAI technologies.

## 1. INTRODUCTION

Browser extensions are software add-ons, typically built with HTML, CSS, and JavaScript, that enhance and customize the web browsing experience. They have become indispensable tools for millions of users, offering functionalities ranging from ad blocking and password management to productivity and accessibility aids. The *Chrome Web Store* (*CWS*), the primary distribution platform for *Chrome* extensions, hosts as many as 180K extensions [1], with typical users installing 8–12 extensions each.

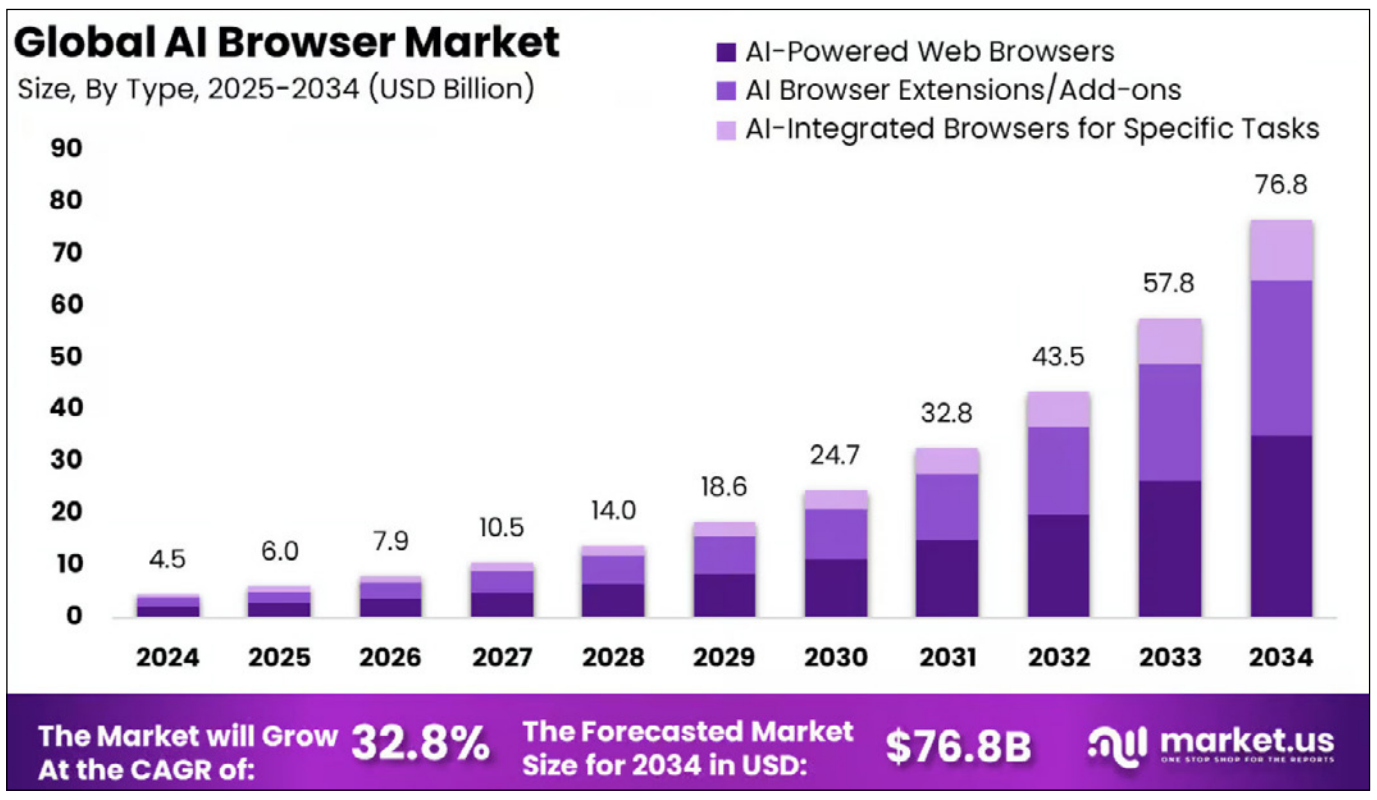

*Figure 1: Global AI browser market (2025–2034) with a 32.8% compound annual growth rate (CAGR); the 'AI Browser Extensions/Add-ons' segment is a major and rapidly growing portion. (Market.us, July 2025 [2]).*

The recent explosion in demand for Generative AI (GenAI) features is driven by users' demand to enhance their productivity. Capabilities like text summarization, content generation, and contextual assistance are a natural fit for the browser extension model, which allows these features to be seamlessly integrated into a user's existing workflow on any website. This trend is reflected in *Google* SEO keyword statistics, which show a 900% year-over-year increase in searches for the keywords 'chrome extension ai'. Furthermore, the AI-Summary subcategory of *Chrome Web Store* extensions has experienced rapid growth in new releases, increasing from 16 in 2023 to 41 extensions year-to-date in 2025, as illustrated in Figure 2. According to *Market.us* [2] the global AI browser market is projected to grow at a compound annual growth rate (CAGR) of 32.8% to reach $76.8 billion by 2034. Significantly, AI browser extensions and add-ons constitute a major and growing segment of this multi-billion dollar market.

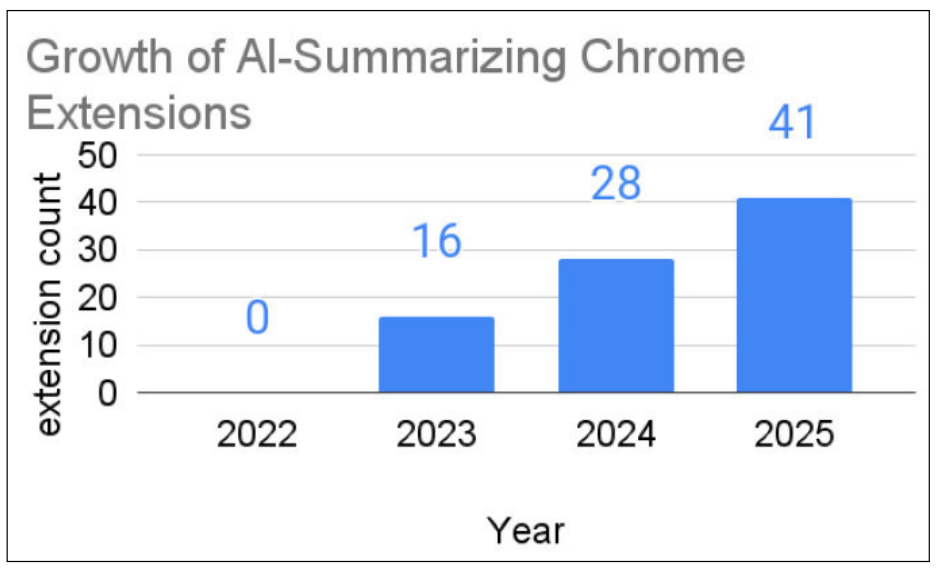

*Figure 2: Growth in the number of available 'AI-Summarizing' Chrome extensions from 2022 to 2025.*

However, the rapid growth and widespread adoption of the *CWS* ecosystem can create opportunities for misuse. Browser extensions typically require access to a variety of browser APIs and user data to provide their advertised functionalities, which, combined with their popularity, can make them attractive targets for malicious actors. This paper examines this evolving threat landscape. The contributions of this paper are as follows:

1. We identify, with quantitative support, that attackers are exploiting the Generative AI trend to distribute malicious browser extensions.
2. We provide an empirical analysis of 5,551 AI-themed extensions launched between 1 January and 15 September 2025, which led to the identification of 154 previously unreported malicious extensions. We incorporate 187 malicious extensions reported in prior public disclosures, resulting in a total of 341 known malicious extensions. Among these, we highlight the 29 that are GenAI-themed, offering insight into this emerging threat category and the tactics used by attackers.
3. We systematically break down the tactics, techniques, and procedures (TTPs) used by GenAI malicious extensions. Case studies highlight techniques such as adversary-in-the-browser, impersonation, prompt hijacking, and bait-and-switch updates, as well as the use of malicious redirection as a tactic to achieve affiliate fraud and PUP delivery.
4. We identify key signals across extension metadata, static code properties, and runtime behaviour, showing how combining these indicators enables effective detection of suspicious extensions.

## 2. BACKGROUND

In this section, we provide background on Chrome extensions, including the fundamental components of Chrome extension architecture (Section 2.1), the security mechanisms that govern it (Section 2.2), and key security changes introduced in the transition from Manifest Version 2 (MV2) to Version 3 (MV3) (Section 2.3).

### 2.1 Chrome extension architecture

A *Chrome* extension is a collection of files and assets, such as HTML, CSS, and JavaScript, that are packaged into a single .crx file for distribution through the *Chrome Web Store* (*CWS*) or external sources. Every extension is assigned a unique 32-character ID derived from the public key of the publisher. This ID ensures the extension can be consistently referenced across different installations and updates. Figure 3, taken from Chrome Developer documentation [3], shows the user interface elements of a *Chrome* extension (CRX).

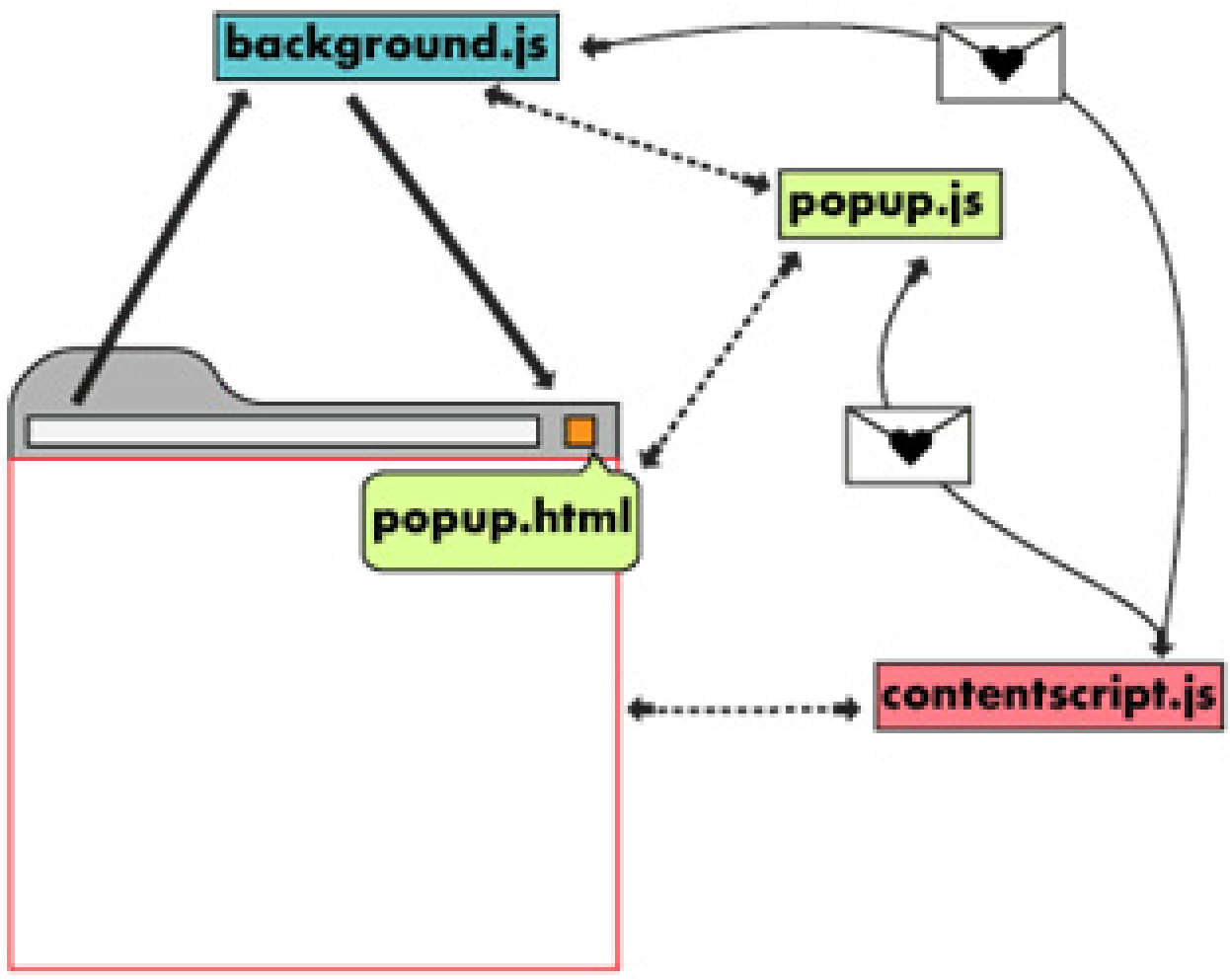

*Figure 3: User interface elements of a Chrome extension (CRX), adapted from the Chrome Developer documentation [3].*

- **Service worker (background script)**: The service worker (or the background script in MV2 [4]) is the extension's privileged core and event-handling centre. Unlike other components, it runs in a browser-level process, not within the context of any single web page. This elevated context grants it access to the full suite of `chrome.*` APIs permitted by the manifest. In MV3, service workers are event-driven, meaning they are initialized to respond to a specific browser event (such as a new tab being created or a message being received) and terminate after a period of inactivity.
- **Content scripts**: Content scripts are JavaScript files that run in a sandboxed environment within web pages. They are injected directly into pages and executed alongside the page's own scripts, with direct access to the Document Object Model (DOM). This enables them to read and modify page content, such as updating text, interacting with forms, and altering the page layout. Content scripts operate in an 'isolated world' and have access to only a limited subset of the chrome.* APIs. To perform actions requiring broader privileges (e.g. cross-origin network requests or access to browser storage), they communicate with the service worker via message passing.

- **Popup (and other UI elements)**: The popup is an HTML page that provides a user-facing interface, displayed when the extension's icon is clicked. While it runs in the extension's context with access to its permitted APIs, it is ephemeral and only active when visible.
- **Manifest file**: This manifest.json file is the foundational blueprint of the extension. It is the first file processed by the browser and serves as a declarative entry point, defining the extension's identity, version, and, most critically, its security boundary. Here we describe the key fields in the manifest file:

1. `manifest_version`: specifies the manifest version the extension targets (e.g. version 2 or 3), which dictates the available APIs and security policies.
2. `permissions`: an array of strings declaring which privileged `chrome.*` APIs the extension requires (e.g. `storage`, `cookies`, and `alarms`).
3. `host_permissions`: an array defining the specific domains or URL patterns (e.g. `https://*.google.com/*` and `<all_urls>`) on which the extension can run its scripts and access data.
4. `background`: declares the service worker script.
5. `content_scripts`: defines which scripts to inject into which web pages.
6. `action`: specifies the popup HTML file and icon.

### 2.2 Key capabilities and security features

The extension components described in Section 2.1 are empowered by a set of core capabilities, and are constrained by specific security features built into the browser, which we detail as follows:

- **Permissions model**: By default, extensions operate in a sandboxed environment with minimal capabilities. To gain additional functionality, they must explicitly declare their required permissions in the manifest.json file. These permissions are presented to the user as a consent mechanism upon installation, outlining the capabilities the extension requires to function. For example, host permissions such as '<all_urls>' allow an extension to interact with any website.
- **Browser extension APIs**: Extensions interact with the browser and modify its behaviour through a suite of JavaScript APIs, accessible via the `chrome.*` namespace (e.g. `chrome.tabs`, `chrome.storage`, `chrome.declarativeNetRequest`). The permissions declared in the manifest file determine which APIs an extension can access, shaping the scope of its functionality.
- **Message passing**: Because components such as content scripts and service workers operate in isolated environments, a dedicated communication protocol is required for them to interact. Message passing enables these components to exchange data in a structured way. The primary methods, `chrome.runtime.sendMessage` and `chrome.tabs.sendMessage`, are fundamental to coordinating complex actions, such as a content script sending information from a web page to a service worker for processing.

### 2.3 Manifest V2 to V3

*Google* is in the process of transitioning browser extensions from Manifest Version 2 (MV2) to Manifest Version 3 (MV3), a shift that began with its initial announcement in 2019 [5]. Support for MV2 in the *Chrome Web Store* was gradually phased down starting in 2022, with full deprecation currently scheduled for 2025 [6]. The key changes include:

- **No remotely hosted code**: MV3 prohibits extensions from executing JavaScript fetched from a remote server. All code must be included in the extension package, making it available for review during the *CWS* vetting process.
- **No dynamic code execution**: MV3 enforces a strict content security policy (CSP) [7] that disallows the use of `eval()`, `new Function()`, and other forms of string-based code execution. This prevents extensions from dynamically assembling and executing JavaScript at runtime. For example, the following code will fail in an MV3 extension:

```
eval("console.log('Hello World')");
```

Attempting this will result in a runtime error in the background service worker:

```
Uncaught EvalError: Code generation from strings disallowed for this context
```

Additionally, *Chrome* will log a console warning:

```
Refused to evaluate a string as JavaScript because 'unsafe-eval' is not an allowed source
of script in the following Content Security Policy.
```

This restriction ensures that all executable code is statically packaged within the extension, making it auditable and preventing last-mile obfuscation attacks.

- **Service workers**: MV3 replaces persistent background pages with event-driven service workers, making it harder for malware to maintain long-term persistence.
- **Declarative net request API**: The transition to Manifest V3 deprecates the powerful but risky blocking version of the `chrome.webRequest` API, which previously allowed extensions to intercept and modify all network traffic in real time. MV3 pushes developers towards the `chrome.declarativeNetRequest` API, where an extension declares rules for blocking or modifying requests in advance, rather than processing them with live code. This limits the extension's ability to read sensitive network data.

Although these changes are designed to make extensions safer, our analysis in Section 5 shows how attackers have adapted, using these security changes for data exfiltration and malicious redirection, bypassing their intended purpose.

## 3. RELATED WORK

The study of malicious browser extensions is a well established field in security research. Contributions range from large-scale measurement studies that characterize the threat landscape to novel methodologies for their detection, analysis, and mitigation. This section provides a survey of key academic works and significant industry threat reports that have informed the understanding of these threats.

Academic research has provided foundational methods for analysis. Measurement papers, such as [1], have demonstrated that malicious and policy-violating extensions can persist in the *Chrome Web Store* for long durations and that many extensions contain known vulnerabilities. [8] pioneered dynamic analysis with 'Hulk', a system using honey pages and fuzzing to elicit malicious behaviour. This line of research, focusing on the observation of runtime behaviour to uncover risks, has since expanded to measure more subtle privacy harms beyond overt malware. For instance, subsequent work has quantified how extensions contribute to browser fingerprinting (XHOUND) [9] and measured how their permissions can be abused to diffuse tracking powers across the web (Extended Tracking Powers) [10]. More recent systems, such as 'Arcanum' [11], build on these principles, introducing modern dynamic taint analysis sandboxes to precisely track how extensions access and exfiltrate sensitive data from web page content. Others have focused on detecting malicious updates by analysing code deltas, a direct response to the 'bait-and-switch' technique [12]. Complementing dynamic approaches, static analysis techniques such as DoubleX [13] detect vulnerable data flows in extensions at scale. In parallel, another line of research has focused on applying machine learning for detection. [14] demonstrated that the sequence of API calls, analysed with a Recurrent Neural Network (RNN), could effectively identify spying extensions, while [15] also used a machine learning approach but focused on feature engineering from static files, and [16] presented a holistic deep learning-based review on detecting malicious browser extensions and links.

Echoing these academic findings, industry reports consistently highlight the vast scale of the problem. A 2025 analysis of over 300K extensions reported by *Dark Reading* concluded that 51% requested overly permissive access, creating a significant security risk [17]. The real-world impact of this was demonstrated by a large-scale spyware campaign disclosed by *Koi Security* and *Malwarebytes*, which affected millions of *Chrome* and *Edge* users through extensions that were harvesting browsing data [18]. A historical precedent for such large-scale data harvesting was the 2019 'DataSpii' leak, where extensions exfiltrated the browsing history of millions of users [19].

A significant body of research and reporting focuses on specific attack vectors. One of the most prominent is supply chain attacks, where legitimate extensions are compromised. A canonical example is The Great Suspender, a popular extension that was sold to a new developer and subsequently updated with malicious code [20]. More recent incidents include the 2022 compromise of the SearchBlox extension to target *Roblox* players [21], the 2024 targeted attack reported by *Sekoia* where spear phishing was used to compromise developer accounts and push malicious updates [22], and the 2025 compromise of the Cyberhaven extension, investigated by *Secure Annex*, which highlighted the risks of even enterprise-focused browser extensions being weaponized [23].

Another prevalent vector is sideloading, where attackers bypass the official *Chrome Web Store* entirely and trick users into installing a malicious extension directly. *Trend Micro* detailed the ParaSiteSnatcher campaign, which used sideloading to target users in Brazil [24]. *Fortinet* reported on RolandSkimmer, where a sideloaded extension was a key component in a credit card theft operation [25]. Similarly, *Netskope* uncovered a campaign delivering LegionLoader via a sideloaded fake *Google Drive* extension [26]. This vector has also been attributed to state-sponsored actors, with North Korean threat groups observed using sideloaded extensions in their attacks [27].

While prior work has extensively examined established attack vectors and the broader threat landscape, this paper contributes by examining the emerging trend of GenAI-themed threats, where attackers adapt known vectors with novel lures to increase effectiveness.

## 4. DATASET AND DETECTION METHODOLOGY

The findings presented in this paper are grounded in the analysis of two datasets: a broad collection of potentially relevant extensions and a curated set of ground-truth confirmed malicious extensions. This section details the composition of these datasets and the methodology used to analyse them.

### 4.1 Dataset curation

The analysis in this work is grounded in a final ground-truth dataset of 341 malicious extensions, which was compiled from two distinct sources.

- **Public disclosures**: We begin by compiling an initial set of 187 confirmed malicious extensions from public disclosures. This set was sourced from a comprehensive review of existing academic literature, recent security blogs, and social media posts from threat researchers, representing the publicly known state of malicious browser extensions.
- **Internal detections**: To investigate the GenAI trend, we initiated our study by compiling a targeted dataset of 5,551 unique *Chrome* extensions released between 1 January and 15 September 2025, which included AI-related keywords in their name or description. By applying the detection methodology detailed in Section 4.2 to this targeted collection, we successfully detected an additional 154 malicious extensions.

Combining our new detections with the publicly disclosed set leaves us with the final ground-truth dataset of 341 known malicious extensions. The results presented in this paper are based on the analysis of this comprehensive set. Of the 154 malicious extensions we detected through our methodology, 29 were specifically GenAI-themed, and these form the basis of the detailed case studies presented in Section 5.

| | | | |
|---|---|---|---|
| pocfdebmmcmfanifcfeeiafokecfkikj | DeepSeek AI \| Free AI Assistant | Public disclosure [30] | 3.0.1 |
| aeibljandkelbcaaemkdnbaacppjdmom | Manus AI \| Free AI Assistant | Public disclosure [30] | 0.1.1 |
| gmbebpcapalekeaoekfhpbioilghcfmp | AI Browser Manager™ | Public disclosure [31] | 1.61.1324.0 |
| oghbffaoaooigagpockijkpfpgmnibkh | ChatGPT Search for Chrome™ | Public disclosure [31] | 1.61.1444.0 |
| cedgndijpacnfbdggppddacngjfdkaca | Wayin AI | Public disclosure [23] | 0.0.11 |
| bbdnohkpnbkdkmnkddobeafboooinpla | Search Copilot AI Assistant for Chrome | Public disclosure [23] | 1.0.1 |
| bibjgkidgpfbblifamdlkdlhgihmfohh | AI Assistant - ChatGPT and Gemini for Chrome | Public disclosure [23] | 0.1.3 |
| befflofjcniongenjmbkgkoljhgliihe | TinaMind - The GPT-4o-powered AI Assis-tant! | Public disclosure [23] | 2.13.0 |
| pkgciiiancapdlpcbppfkmeaieppikkk | Bard AI chat | Public disclosure [23] | 1.3.7 |
| epikoohpebngmakjinphfiagogjcnddm | AI Shop Buddy | Public disclosure [23] | 2.7.3 |
| bgejafhieobnfpjlpcjjggoboebonfcg | ChatGPT Assistant - Smart Search | Public disclosure [23] | 1.1.1 |
| epdjhgbipjpbbhoccdeipghoihibnfja | GPT 4 Summary with OpenAI | Public disclosure [23] | 1.4 |
| lbneaaedflankmgmfbmaplggbmjjmbae | ChatGPT App | Public disclosure [23] | 1.3.8 |
| hmiaoahjllhfgebflooeeefeiafpkfde | Hi AI | Public disclosure [23] | 1.0.0 |
| jmpcodajbcpgkebjipbmjdoboehfiddd | DeepSeek AI Chat | Public disclosure [32] | 2.4 |
| ihdnbohcfnegemgomjcpckmpnkdgopon | AI Sentence Rewriter | Public disclosure [32] | 2.1 |
| mldeggofnfaiinachdeidpecmflffoam | AI Writer | Public disclosure [32] | 2.1 |
| pndmbpnfolikhfnfnkmjkkpcgkmaibec | AI Ad Generator | Public disclosure [32] | 1.4 |
| eebihiecleccoidddmjcencomodomdoei | Supersonic AI | Our work [33] | 1.0.6 |
| dcjfbgppfdokmjgajnnkgdmkdeiloigh | Picsart: AI Photo Video Editor | Our work [34] | 2.4.0 |
| khggmpoiighnlkmppaofcklcpiclaelb | Photoroom AI Photo Editor | Our work [34] | 1.0 |
| jhhjbaicgmecddbaobeobkikgmfffaeg | Chat AI for Chrome | Our work [35] | 1.1.2 |
| bpeheoocinjpbchkmddjdaiafjkgdgoi | Chatgpt for chrome, AI extension for Chrome | Our work [35] | 1.0.0 |
| pjcfmnfappcoomegbhlaahhddnhnapeb | Meta Llama search | Our work [35] | 1.0 |
| ecimcibolpbgimkehmclafnifblhmkkb | Perplexity search | Our work [35] | 1.0 |
| akfnjopjnnemejchppfpomhnejoiiini | Claude search | Our work [35] | 1.1 |
| jijilhfkldabicahgkmgjgladmggnkpb | GenAISearch | Our work [35] | 1.1 |
| lnjebiohklcphainmilcdoakkbjlkdpn | Chatgpt search | Our work [35] | 1.0 |
| boofekcjiojcpcehaldjhjfhcienopme | AI ChatGPT | Our work [35] | 1.0.3 |
| mkhhkiopdkhhelagdjpknbemdeonhhho | QuizFlash | Our work | 1.0 |

*Table 1: List of GenAI-themed malicious extensions reported.*

### 4.2 Detection methodology

To identify malicious candidates within our initial dataset of 5,551 extensions, our approach relies on a multi-signal filtering approach that correlates a wide array of signals. Our approach is designed to identify all ground-truth malicious extensions and produce candidates from the initial dataset for manual review and confirmation.

Table 2 summarizes the key signals guiding our investigation, grouped into metadata, static analysis, and runtime (behavioural and network) analysis.

| Category | Signal | Description |
|---|---|---|
| Extension metadata | Author Reputation | Evaluate the trustworthiness of the extension author based on past extensions, known associations, or recent accounts. |
| | Extension Description | Check for misleading, vague, or overly generic descriptions that could conceal malicious intent. |
| | User Install Count or Low Review Count | Low install counts or sudden spikes may indicate suspicious new or repackaged extensions. |
| | Extension Age | Newly published extensions or recently updated extensions may be more likely to contain malicious updates. |
| Source code analysis | Obfuscated Code | Use of packing, string encoding (e.g. Base64), or convoluted control flows to hide malicious logic from static analysis and review. |
| | Risky API Usage | Invocation of powerful APIs, such as `declarativeNetRequest` for traffic interception or `scripting.executeScript` for arbitrary code execution on pages. |
| | Remote Code Fetching | Dynamically fetching and executing JavaScript from an external, attacker-controlled server, violating store policy and enabling bait-and-switch attacks. |
| Network intelligence | Suspicious Domains | Communication with newly registered domains (NRDs), domains with poor reputation, or those present on threat intelligence blocklists. |
| | C2 Communication | A pattern of beaconing to a command-and-control (C2) server to receive new instructions, update configurations, or exfiltrate data. |
| Behavioural monitoring | Unauthorized DOM Manipulation | Modification of web page content (the DOM) to inject advertisements, phishing forms, keyloggers, or cryptocurrency miners without user consent. |
| | Traffic Redirection | Hijacking user navigation to fraudulent affiliate links, PUP (Potentially Unwanted Program) download sites, or phishing pages. |
| | Data Exfiltration | Capturing and sending sensitive user data (such as credentials, PII, or session cookies) to a remote endpoint controlled by the attacker. |

*Table 2: Key signals for detecting malicious browser extensions.*

- **Extension metadata analysis**: We collect extension metadata from *Chrome-Stats* [28], a third-party service that aggregates and tracks metadata from the *Chrome Web Store*. The analysis of an extension's metadata provides important contextual clues about its reputation and potential risk. The author's reputation is evaluated based on whether they have previously published extensions that were flagged as malware, left unmaintained, or removed for policy violations. We consider the age of an extension as another key factor in evaluating its risk profile. We also flag extensions that have low user count or show signs of artificially generated reviews, as signals of possible malicious intent.
- **Static code analysis**: Static analysis examines the source code of an extension without executing it. This process can reveal signals of its potential capabilities and uncover deliberate attempts at evasion. We focus on identifying obfuscated or heavily minified code, complex string manipulations, or convoluted control flows designed to hide malicious logic from static scanners and human reviewers. A critical signal is the presence of any mechanism for remote code fetching, which enables attackers to execute remotely hosted JavaScript that is a direct violation of Manifest V3 policy. We also identify risky API usage, such as the invocation of powerful functions like `chrome.scripting.executeScript` for arbitrary code execution or `declarativeNetRequest` for traffic interception.
- **Behavioural and network intelligence**: Dynamic analysis is essential for observing runtime behaviour and uncovering threats that are not visible through static inspection. This involves monitoring network traffic and in-browser actions. We monitor communication with suspicious domains, particularly newly registered domains (NRDs) and those flagged as low reputation by third-party sources such as *VirusTotal* [29] or internal (*Palo Alto Networks*) intelligence. This also includes identifying patterns of C2 communication, such as regular beaconing to a command-and-control server to exfiltrate data or receive new instructions. On the behavioural side, we watch for unauthorized DOM manipulation, where the extension injects advertisements, phishing links, or keyloggers into a web page without user consent. Traffic redirection to fraudulent affiliate links or PUP download sites is another critical indicator of malicious activity. Finally, we explicitly monitor for data exfiltration, where sensitive information – such as credentials, PII, or session cookies – is captured from the browser and sent to an attacker-controlled remote endpoint.

Our current triaging process relies on manual correlation and expert human review. However, these signals provide a clear foundation for automation. In the future, we plan to explore similar trends in other browser platforms such as *Edge* and *Firefox*.

## 5. DEEP DIVE INTO ATTACKER TACTICS AND TECHNIQUES

This section provides a detailed analysis of the malicious activities observed across 29 GenAI-themed extensions (Table 1). We examine the specific tactics, techniques, and procedures (TTPs) used by threat actors, presenting hands-on case studies based on our analysis. We observe that the threats are mainly associated with two objectives: data exfiltration and malicious redirection.

### 5.1 Data exfiltration

These attacks aim to collect sensitive user information from the browser, exploiting extension permissions to access data without user consent. We observed three prevalent techniques, described as in the following subsections.

#### *5.1.1 Adversary-in-the-browser*

In the adversary-in-the-browser (AiTB) technique, a malicious extension leverages its permissions to extract sensitive data directly from the Document Object Model (DOM) of a trusted web application. Operating within the user's browser, this privileged position enables access to information that would normally remain within the trusted boundary of the application.

**Lure**: The attack is demonstrated by the Supersonic AI (eebihieclccoidddmjcencomodomdoei) extension (Figure 4), listed on the *Chrome Web Store* as an AI-powered email assistant for *Gmail* and *Outlook*. According to its listing, it helps users 'reply faster, write smarter, and stay on top of what matters', offering features such as one-click AI-generated replies and email summaries. As per its description, the extension targets busy professionals. It displayed a 5/5 star rating and has been installed by 26 users. Human review indicates that these reviews could be artificially generated.

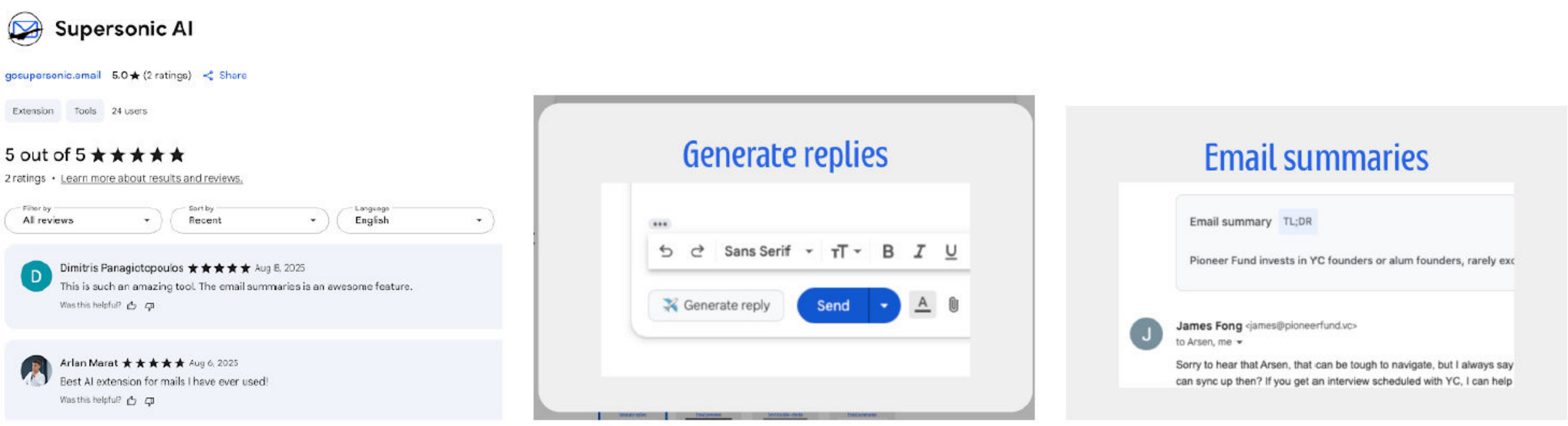

(a) Listing page (b) Generate reply feature (c) Email summary feature

*Figure 4: Chrome Web Store listing and features of the Supersonic AI extension, an example of an adversary-in-the-browser (AiTB) technique.*

##### *Attack flow 1: active data exfiltration via feature interaction:*

The primary data exfiltration occurs through the extension's advertised features.

- **Scenario**: The attack is triggered when a user interacts with the *Gmail* or *Outlook Web* page. For example, in a corporate environment, a user may receive an email containing confidential data, such as M&A term sheets or personally identifiable information (PII).
- **User interaction**: The user invokes the extension's 'Generate reply' feature, which produces a context-aware, professionally worded response, masking its background activity.
- **Technical analysis**: Code and network analysis reveal a multi-step exfiltration process:
    1. **Content script (`content.js`) data harvesting**: An event listener bound to the 'Generate reply' button activates the content script injected into the webmail client's DOM. This script programmatically accesses the content of the entire email thread, including the sensitive message.

```
const response = await new Promise((resolve,
,→ reject) => {
   chrome.runtime.sendMessage({
     action: 'generateReply',
     data: {
       email_content: lastEmailContent,
       context: 'Reply to the last email
       '→ in this thread',
       chat_history: chatHistory
       }
     });
});
```

2. **Message passing**: The harvested data is packaged into a JavaScript object. To transfer this data from the sandboxed content script to the more privileged background script, the extension utilizes the `chrome.runtime.sendMessage` API. The architecture follows a common design pattern, which aligns with typical extension behaviour.

3. **Service worker (`background.js`) exfiltration**: The background service worker is configured with a listener for the `generateReply` message. Upon receipt, it executes a `fetch` request, sending the data payload to an external untrusted endpoint.

```
async function generateReply(requestData) {
   const response = await
   '→ fetch(CONFIG.ENDPOINTS.GENERATE_REPLY,
   '→ {
     method: 'POST',
     headers: {
        'Content-Type':
        ,→ 'application/json',
        'Authorization': 'Token
        ,→ ${token}',
     },
     body: JSON.stringify(requestData) //
     '→ Exfiltration payload
   });
}
```

4. **Confirmation**: Dynamic analysis captured this behaviour, logging a `POST` request to `hxxps://api[.]gosupersonic[.]email/api/generate-reply/`. The request body contained the full, unredacted content of the confidential email, confirming the exfiltration of sensitive data.

#### *Attack flow 2: passive data exfiltration without direct interaction*

A more notable attack flow involves data exfiltration that occurs without any direct user interaction with the extension's features.

- **Scenario**: A user simply opens an email. This email could contain sensitive tokens, such as a password reset PIN.
- **Technical analysis**: The extension's content script is programmed to access the content of any email displayed in the DOM. This data is subsequently transmitted as part of the 'Email summaries' feature. Network analysis as shown in Figure 5 captured the outgoing `POST` request to the `generate-summary/` endpoint containing the following payload:

```
{
  "content": "Verify your LinkedIn account with
  ,→ code 758643...",
  "created_at": "2025-09-17T21:42:51.572201",
  "summary_type": "tldr"
}
```

- **Impact**: The attacker obtains the password reset PIN '758643' in near real time, granting them control over the user's corresponding account. This passive mechanism allows data to be stolen from any viewed email, regardless of whether the user actively interacts with the extension.

- **Persistence and evasion**: Despite the *Chrome Web Store*'s strict policies, the extension remains available. Analysis of its associated domain (`api[.]gosupersonic[.]email`) on *VirusTotal* returned a detection rate of 1/97, reflecting limited coverage by existing blocklists.

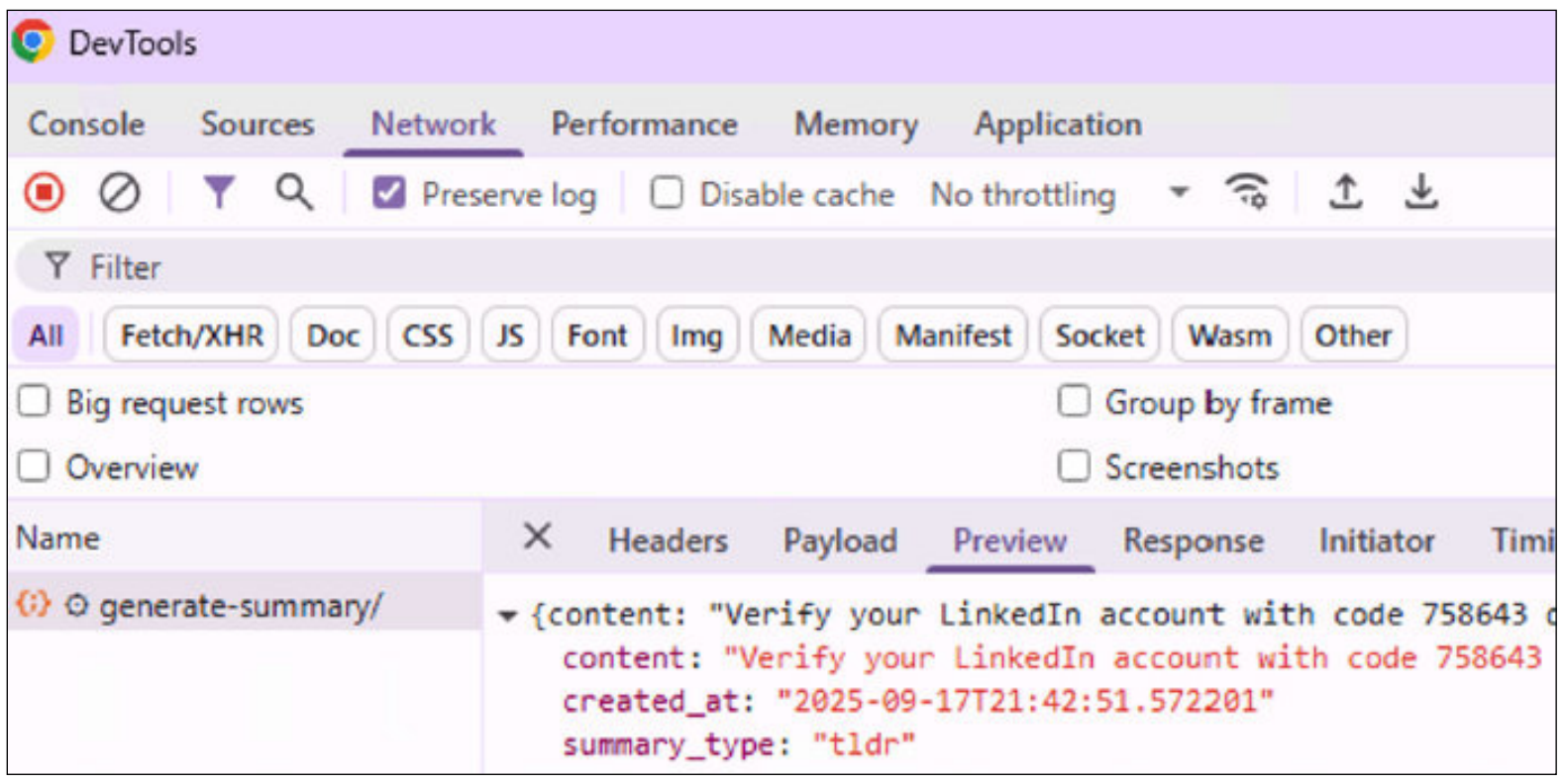

*Figure 5: Supersonic AI extension: LinkedIn OTP data exfiltration as seen in sandbox network logs.*

### *5.1.2 Impersonation, dual functionality, and bait-and-switch*

We analyse a set of impersonating extensions, focusing on DeepSeek AI | Free AI Assistant (pocfdebmmcmfanifcfeeiafokecfkikj) as a representative example. Initial intelligence from *DomainTools* [30] in May 2025 identifies this family of extensions and highlights their use of impersonation and dual functionality. Our research builds on these findings by conducting a version-by-version analysis of the extensions' lifecycle, which reveals a third technique employed by the attackers: bait-and-switch, as shown in Figure 6c and the technical analysis section below.

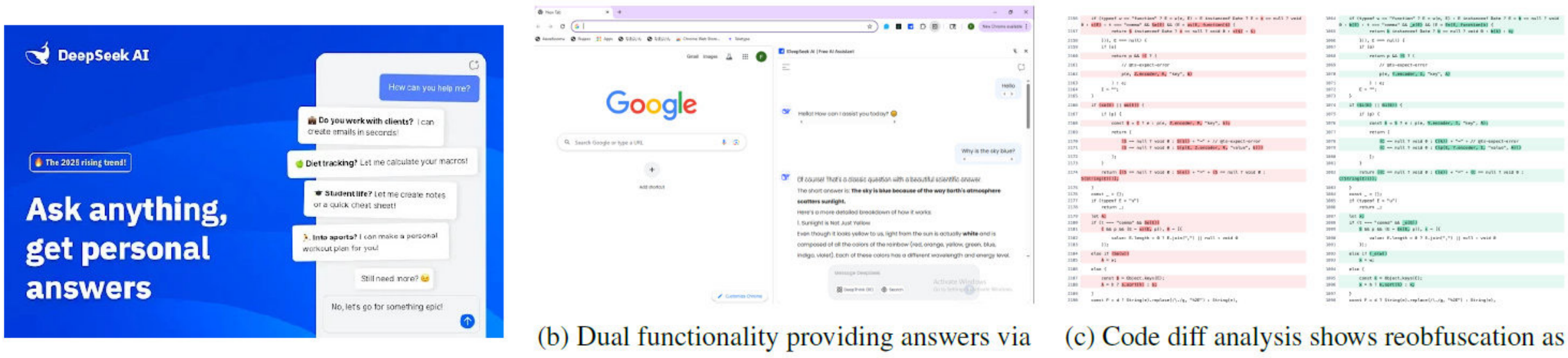

(a) Chrome Web Store listing modern UI.

(b) Dual functionality providing answers via the legitimate `api[.]deepseek[.]com`.

(c) Code diff analysis shows reobfuscation as a part of the bait-and-switch technique.

*Figure 6: Overview of the DeepSeek AI extension across its UI, functionality, and code evolution.*

- **Impersonation**: The extension DeepSeek AI | Free AI Assistant impersonates legitimate AI service `deepseek[.]com`. The extension replicates the UI, logos, and overall design of the legitimate product, as seen in Figure 6a. It had accumulated approximately 350 installs, indicating that the impersonation technique can successfully attract users.
- **Dual functionality**: This technique by this extension was first reported in [30]. The extension provides the advertised functionality by forwarding user queries to the legitimate backend (`api[.]deepseek[.]com`) and displays the corresponding responses as seen in Figure 6b. Network analysis indicates that, alongside this legitimate traffic, the extension also sends `GET` requests to a known malicious endpoint (`https://api[.]glimmerbloop[.]top`), which has 11 *VirusTotal* detections at the time of writing this paper. This parallel channel is used for C2 communication and other extension operations.
- **Bait-and-switch update**: Our analysis of the update history of the extensions reveals an additional finding: the use of a switch-and-bait update. This evasive technique may have helped bypass the *Chrome* extension review process and increase user trust. The different versions of the extension codebase for this historical analysis were obtained from *Chrome-Stats* [28]. We found that DeepSeek AI | Free AI Assistant was initially published with benign versions (0.0.1 and 0.0.2). These early versions did not contain malicious code. A subsequent update (version 3.0.1) introduced additional functionality that included malicious behaviour.
- **Technical analysis**: A differential analysis was conducted between the benign (v0.0.2) and updated (v3.0.1) versions of the extension. The analysis showed that the large 8,000-line background script had been re-obfuscated. This reobfuscation resulted in a large and noisy code differential, making it less straightforward to identify changes through manual or automated review. In addition, a configuration object was added that contains the attacker's malicious endpoint.

```
var yt = {
    CEB_ENCRYPT_LOCAL_STORAGE: "true",
    CEB_BASE_URL:
    ,→ "https://api.glimmerbloop.top/api", //
    ,→ Malicious Endpoint Added
    CEB_STARTUP_DELAY: "15",
    CEB_ACTIVITY_TIMEOUT: "5",
    CEB_REDIRECT_URL: "https://www.google.com",
    CEB_INSTALL_ENDPOINT: "/v2/install",
    CEB_NODE_ENV: "production"
}
```

- **Impact**: The layering of these techniques allows the attacker to maintain persistence. For example, the Manus AI | Free AI Assistant extension (aeibljandkelbcaaemkdnbaacppjdmom) extension from the same campaign that leveraged the techniques described in this section remained in the *Chrome Web Store* for approximately three months (March to June 2025).

### *5.1.3 Query hijacking: from search keywords to sensitive prompts*

Query hijacking is a specific technique for data exfiltration that involves programmatically altering browser search settings to capture the user's query input. A common mechanism for this is the `chrome_settings_overrides` key in the manifest, a *Chrome* setting with legitimate applications. For example, the official extension from Perplexity AI uses this setting to properly integrate its service into the browser's omnibox, providing a seamless user experience.

However, our analysis shows that threat actors abuse this same functionality through a combination of impersonation and covert redirection. They create impersonating extensions that mimic legitimate services, use the `chrome_settings_overrides` setting to intercept user queries, and then route this traffic through attacker-controlled servers before forwarding it to the intended destination. This allows them to steal user queries without disrupting the user's workflow.

#### *Known threat: traditional search hijacking*

This attack is demonstrated using the Chat AI for *Chrome* (jhhjbaicgmecddbaobeobkikgmfffaeg), shown in Figure 7a, an extension that claims to enhance browsing by providing AI-powered answers directly in the search bar via `chatgptforchro me[.]com`. The search hijacking technique is well documented and has been observed frequently [36] in web security contexts to capture user search queries. In this extension, the `chrome_settings_overrides` setting is configured to redirect all user searches to a malicious domain (`chatgptforchrome[.]com`), allowing the attacker to harvest search terms.

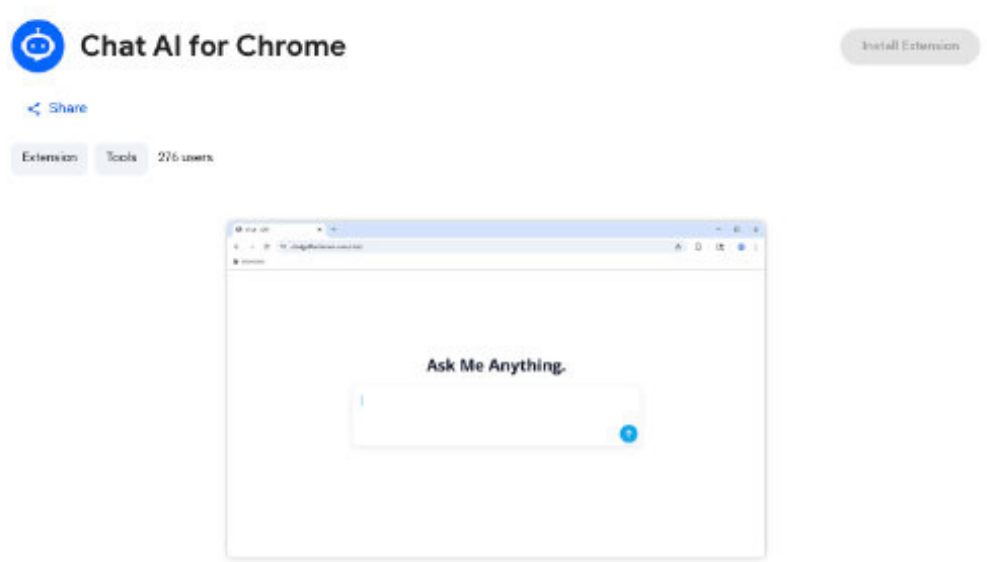

(a) Chat AI for Chrome

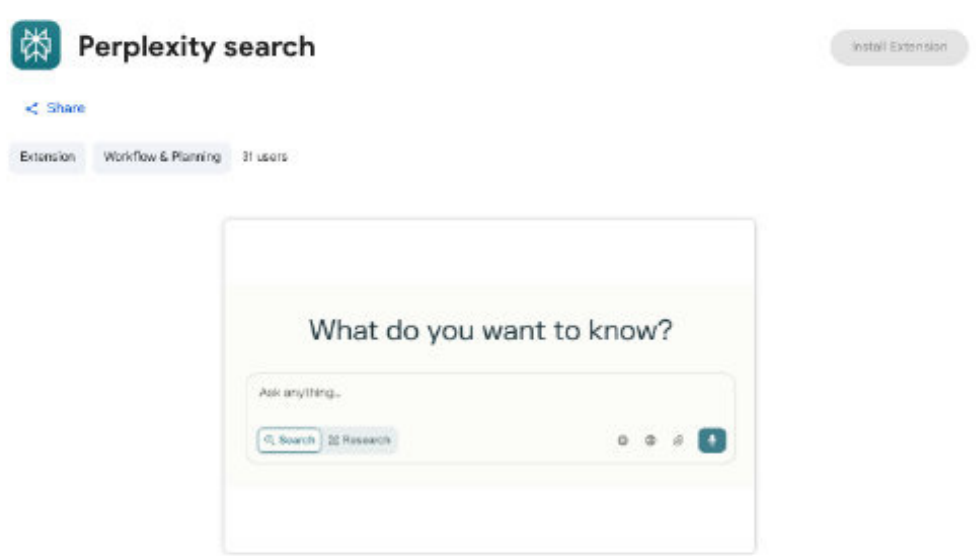

(b) Perplexity Search

*Figure 7: Query hijacking examples: Chat AI for Chrome (searches via chatgptforchrome[.]com to Yahoo) and Perplexity Search (LLM prompts via dinershtein[.]com to Perplexity).*

#### *Evolving threat: prompt hijacking*

The Perplexity Search (ecimcibolpbgimkehmclafnifblhmkkb) extension, shown in Figure 7b, illustrates prompt hijacking, a potentially lucrative evolution of the attack, adapted for the era of conversational AI. As users move from entering simple keywords to submitting more detailed, context-rich prompts to LLMs, the information contained in intercepted queries may change accordingly. These prompts may contain sensitive information, increasing the risk of interception.

- **Attack flow**: By impersonating the legitimate service through its name, the extension tricks users into installing it. It then uses the `chrome_settings_overrides` setting not to connect directly to Perplexity, but to initiate a malicious redirection chain whenever a user enters a prompt in the omnibox.
- **Technical analysis**: Sandbox network logs revealed a two-step process:
    1. **Interception and exfiltration**: The browser is first forced to make a `GET` request to an attacker-controlled server at `https://dinershtein[.]com/perplexity.html?q={UserPrompt}`. This server captures the user's prompt.
    2. **Transparent redirection**: After logging the data, the attacker's server issues a 301 redirect, seamlessly forwarding the user's browser to the legitimate `https://perplexity.ai/search/...` to fetch the AI-generated answer.
- **Impact**: User interaction remains uninterrupted, as the user receives the expected response from Perplexity and is unaware of the interception. Although the redirection resembles traditional search hijacking, the exfiltrated data now consists of conversational prompts rather than simple keywords. As a result, the nature of the captured information differs from that in conventional search hijacking.

### 5.2 Malicious redirection

These attacks aim to benefit the attacker through outcomes such as monetization schemes (e.g. affiliate fraud) or the delivery of potentially unwanted programs (PUPs) by leveraging an extension's privileged position to redirect the user requests.

#### *5.2.1 Affiliate fraud via malicious redirection*

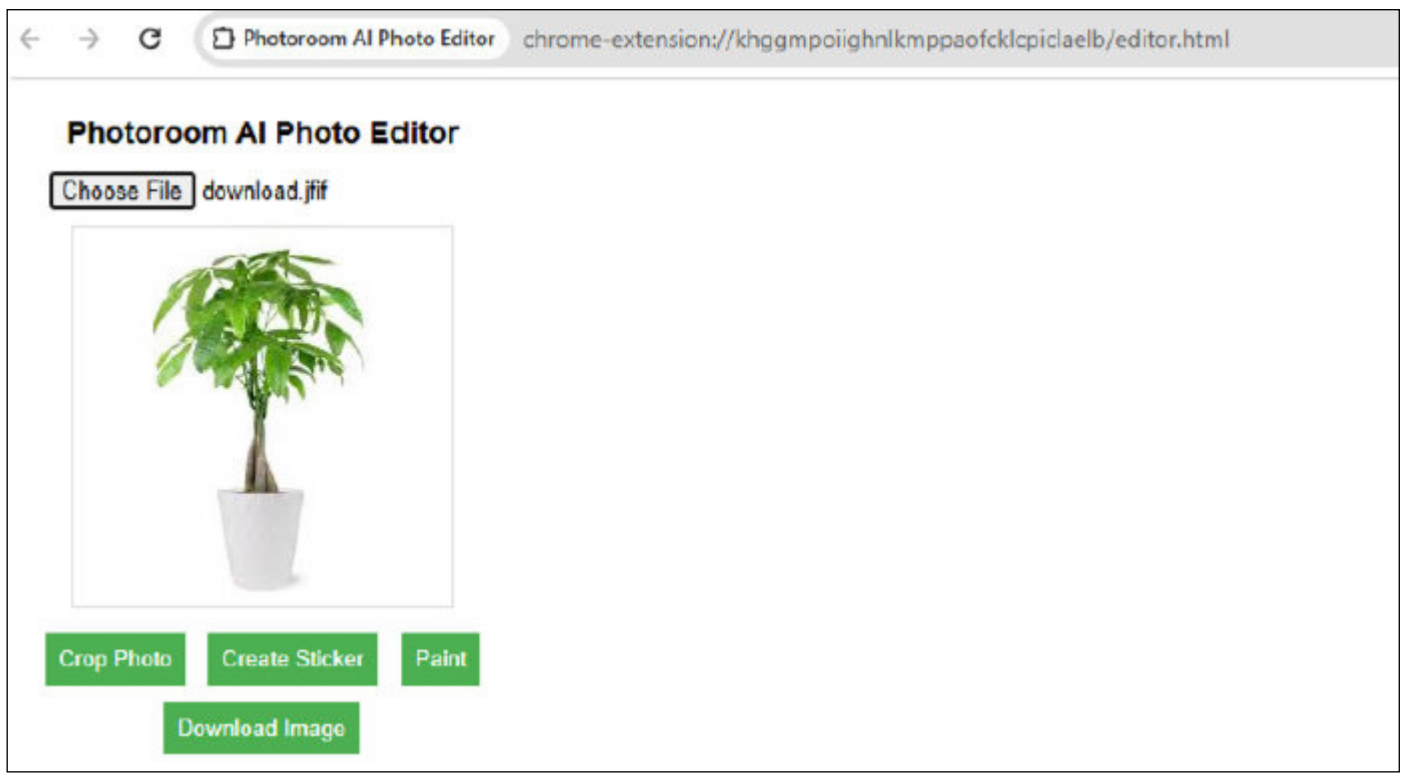

*Figure 8: User interface of the malicious Photoroom AI Photo Editor extension. The minimal features suggest a potential lure, while the extension is positioned to abuse the onInstalled event for affiliate fraud.*

We use the Photoroom AI Photo Editor (khggmpoiighnlkmppaofcklcpiclaelb) extension, shown in Figure 8 as an illustrative case. This technique uses a simple lure and a common browser event to generate fraudulent affiliate revenue.

- **Lure**: The Photoroom AI Photo Editor extension provided minimal functionality but capitalized on the brand recognition of the legitimate 'Photoroom' application.
- **Technical analysis**: The attacker achieves redirection by abusing the `chrome.runtime.onInstalled` event, which invokes a callback immediately after installation.
    1. **Background script trigger**: The background script contains a listener for this event, which programmatically creates a new tab.

```
// Snippet from background.js
chrome.runtime.onInstalled.addListener(() =>
,→ {
    chrome.tabs.create({
      url: "photoroomeditor.html"
    });
});
```

2. **Redirect via iframe**: The local `photoroomeditor.html` page is not the final destination. It contains an iframe that loads content from an attacker-controlled domain (`http://photor-extens[.]uno/`). The content loaded from the attacker's domain is shown in Figure 9.

```
<body>
   <iframe src='http://photor-extens.uno/'
   ,→ allowfullscreen></iframe>
</body>
```

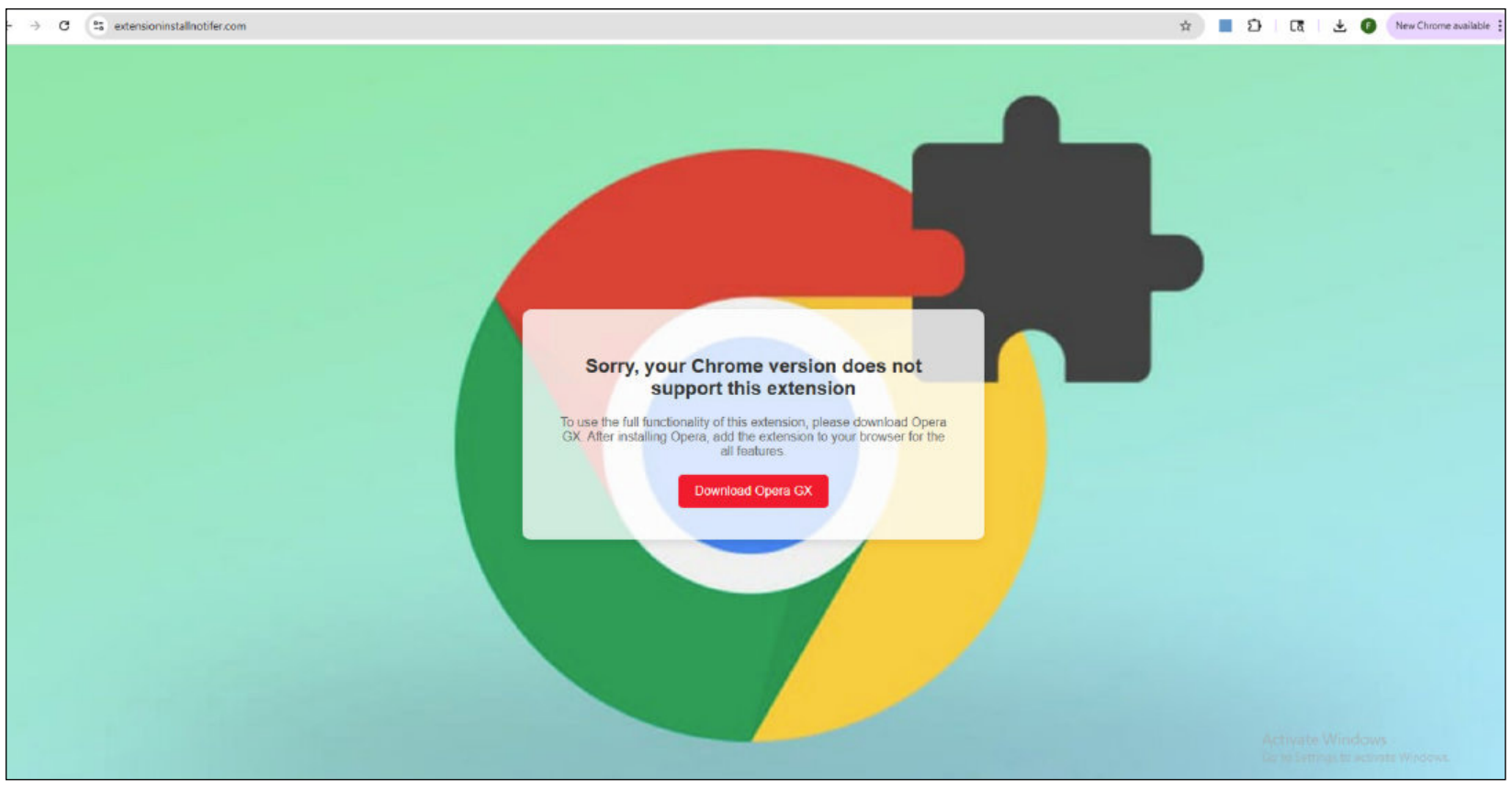

*Figure 9: Deceptive error message prompting the user to click the 'Download Opera GX' button, which is monetized through an affiliate paid acquisition (PA) scheme.*

- **Impact**: The impact of this redirection is twofold. Firstly, the user is subjected to a deceptive social engineering technique. Upon extension installation, they are immediately redirected to a page displaying a deceptive error message: '`Sorry, your Chrome version does not support this extension.`' This message socially engineers them into clicking a prominent '`Download Opera GX`' button as seen in Figure 9. Secondly, this deception is directly monetized through a specific affiliate fraud scheme. Analysis of the link behind the 'Download' button reveals a standard affiliate-tracking URL structured with UTM parameters, such as: `...&utm_medium=pa&utm_source=PWNgames...`. The parameter `utm_medium=pa` explicitly indicates that this is a 'paid acquisition' (PA) marketing campaign. Under this model, the attacker acts as a fraudulent publisher. For each user who is successfully redirected through this unique link and proceeds to install the promoted software (Opera GX), the attacker receives a financial commission from the software vendor's advertising budget.
- **Campaign discovery**: Our analysis revealed that this extension was part of a broader campaign involving at least 68 similar malicious extensions, indicating that the scheme was being operated at scale. This finding was publicly reported through *Palo Alto Networks*' *Unit 42* channel [34].

### *5.2.2 Potentially unwanted program (PUP) delivery via malicious redirection*

We present The Pokemon Cursor ★ Custom Cursor for Chrome™ (gpacldldkpfobgbdabaollodfoilfela) extension as a representative case study. This case study, while not GenAI-themed, comes from our detections encountered while evaluating the GenAI dataset. It demonstrates a common technique where an extension acts as a gateway for more invasive software.

- **Lure**: The Pokemon Cursor ★ Custom Cursor for Chrome™ extension appeals to users by offering custom-themed browser cursors.
- **Technical analysis**: The extension is designed with a component that serves as a lure, redirecting the user toward additional features that are hosted on the attacker's domain.

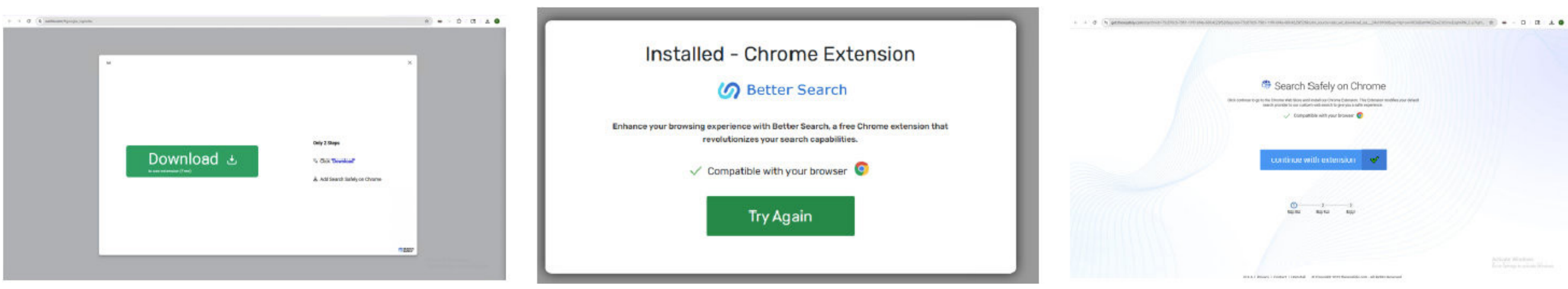

*Figure 10: Multi-stage redirection funnel in the Pokemon Cursor PUP attack, from the deceptive gateway (owhit[.]com) to the intermediary pop-up and finally the PUP landing page (get[.]theresafely[.]com), using social engineering to drive installation.*

1. **Initial vector**: The extension's popup UI includes a 'Get More Cursors' button.
2. **Redirect via button click event**: Analysis of the extension's `cursorpopup.html` file reveals this button is a hyperlink to a grayware domain `owhit[.]com`, as seen below. This domain serves as an entry point to a PUP delivery network. It redirects the user, as seen in Figure 10, to `get[.]theresafely[.]com`, a landing page promoting the Better Search extension, a suspicious PUP that has been reported to perform ad injection and search hijacking.

```
<a href='hxxps://owhit[.]com/'
,→class="action-btn pink"
,→target="_blank">Get More Cursors</a>
```

- **Impact**: The initial, seemingly harmless cursor extension functions as a dropper for more malicious software. Users seeking a cosmetic change are deceived into installing a program that fundamentally compromises their browser's security and integrity. This illustrates how attackers leverage benign functionalities to escalate their access and impact.
- **Campaign discovery**: This PUP delivery was part of a larger campaign involving 75 malicious extensions. This was publicly reported via *Palo Alto Networks*' *Unit 42* channel [37].

## 6. CONCLUSION

Attackers have adapted to the security enhancements of Manifest V3, misusing extension APIs and settings to achieve their objectives. The explosive growth in Generative AI (GenAI) has created a shift in the threat landscape of browser extensions. Our research demonstrates cases where malicious *Chrome* extensions are successfully exploiting this trend, evidenced by statistics such as user installation count and the volume of new GenAI-themed malicious variants discovered. A combination of manifest, static, and dynamic analysis is necessary for the effective detection of these malicious extensions.

## 7. ACKNOWLEDGEMENTS

This work would not have been possible without the invaluable support of our teammates at Internet Security Research, *Palo Alto Networks*: Oleksii Starov, Qinge Xie, Fang Liu, Shehroze Farooqi, Shawn Huang, Xinran Zhang, Jingwei Fan and Yulei Liu.